

\input amstex
\documentstyle{amsppt}
\nologo
\magnification=\magstep1
\pagewidth{13.6cm}
\pageheight{19.0cm}
\hcorrection{-3.2mm}
\baselineskip=22pt
\parskip=5pt
\parindent=8mm
\document
\TagsOnRight
\LimitsOnInts

\def\ref(#1){$^{#1}$}
\font\bigfc=cmbx10 scaled 1200


{\bigfc{\centerline{An approach to quantum gravity }
\centerline{from 4-$\epsilon$ dimension}}}
\vskip 2.0 true cm
\centerline{Kazuo \ GHOROKU${}^1$ \ } \par
\smallskip
\centerline{The Niels Bohr Institute}\par
\centerline{Blegdamsvej 17, DK-2100 Copenhagen \O, Denmark} \par
\bigskip
\vskip 1.0 true cm

\vskip 1.0 true cm
\noindent {\bf{Abstract:}} A calculational scheme of quantum-gravitational
effects on the physical quantities is proposed. The calculations are
performed in 4-$\epsilon$ dimension with $1/N$-expansion scheme, where the
Einstein gravity is renormalizable and it has an ultraviolet fixed-point
within the 1/N-expansion.
In order to perform a consistent perturbation in $4-\epsilon$ dimension,
spin-3/2 fields should be adopted as the N matter-fields whose
loop-corrections are included in the effective action.
After calculating the physical quantities at $4-\epsilon$ dimension,
the four-dimensional aspects of them can be seen by taking the limit
of $\epsilon=0$. In taking this limit,
any higher derivative terms are not introduced as the counter terms since no
divergence appears at $\epsilon=0$ in our scheme.
According to this approach, we have examined the effective potential
of a scalar field to see the possibility of the spontaneous symmetry
breaking due to the
gravitational loop corrections. \par
\noindent -----------------------------------------------------------
\par\noindent ${}^1$ the permanent address:
Department of Physics,
Fukuoka Institute of Technology,\par
\noindent Wajiro, Higashi-ku, Fukuoka 811-02, Japan

\vfill\eject
\smallskip
\noindent{\bf 1. Introduction}
\smallskip
It is an important problem to find a theory which could provide us a
reliable calculation of a quantum gravitational-effects on the physical
quantities. Although superstring theory can be considered as such
a theory, it is hard to get
an available four-dimensional theory of quantum-gravity from it.
Any kind of successful field-theory at four-dimension is renormalizable, so
one would expect such a theory also in the gravity. It is certainly possible
to make a renormalizable, gravitational theory [1] if
we add quadratic terms of the curvature to the Einstein gravity, but
it is not unitary since a massive ghost
appears.
 In order to avoid this ghost, Tomboulis [2] has previously proposed
a 1/N-expansion formalism, where the effective lagrangian ($L_{eff}^{1/2}$)
obtained after integrating out N fermion-fields is used for the perturbative
calculation of the metric-fluctuations.
In this perturbative scheme,
a complex conjugate poles appear instead of the massive ghost,
but the perturbative unitarity can be realized due to the Lee-Wick mechanism
[3]. However the original theory should include higher derivative terms in
order to make the theory being renormalizable, so the massive ghost has not
been removed from the theory. In other words,the problem of the ghost
is not still resolved
even if we hide it behind the complex poles by changing the method of
summing
the perturbative terms through the 1/N-expansion scheme. \par
 On the other hand, the higher derivative terms are not needed in the
effective lagrangian if the 1/N-expansion formalism is applied to the theory
defined at $d=4-\epsilon$ since the corresponding divergence does not
appear in the matter loop corrections
for positive $\epsilon$. Then, it is possible at $d=4-\epsilon$ to give a
renormalizable scheme in which the ghost is absent,
and the unitarity would be retained in this sense.
Several years ago, Smolin [4] has investigated the Einstein gravity
in the dimension less than four in terms of the $1/N$-expansion to show the
existence of an ultraviolet
fixed point.
This implies the possibility of
performing a reliable perturbations at $d=4-\epsilon$ within the 1/N-expansion,
but
such perturbative calculation has not been executed.
Insteadly, he considered the four-dimensional limit of the effective
lagrangian itself, and he arrived at
the usual higher derivative theory by introducing the curvature squared terms
which were necessary to subtract the divergent term at $d=4$. \par
 Here, we propose a different approach to the four dimensional limit
from the theory given at $d=4-\epsilon$ without introducing the higher
derivative terms; (i) Firstly, apply the 1/N-expansion formalism to the
Einstein gravity defined at $d=4-\epsilon$ in
order to resolve the dilemma of the renormalizability and the unitarity, then
calculate the physical quantities at this dimension.
(ii) Nextly, take the limit of $\epsilon=0$ for the calculated quantities
in order to see their physical aspects in the realistic four dimension after
finishing the calculation at $d=4-\epsilon$. \par
 The effective lagrangian used in the step (i) should be obtained by
integrating out the N matter fields with a general-coordinate invariant
regularization. Since a non-invariant regularization was used in ref.[4], we
here recalculated the loop corrections of N fermion fields. But its result
was not used in our effective lagrangian at $d=4-\epsilon$ since
the effective action obtained by adding the loop-correction of N spin-1/2
fields is not available for executing the perturbative calculation at
$d=4-\epsilon$
because of a tachyonic pole in the spin-two field propagator. This
undesirable situation is improved by adopting the spin-3/2 fields as the N
matter-fields instead of the spin-1/2 one.
In this case,
the tachyonic pole disappears as shown in the section 4.
We insteadly find a pair of
complex conjugate poles other than the massless graviton pole.
\par
  The appearance of the unstable and unphysical complex conjugate poles
is inevitable as far as we consider the propagator which could make
the theory being renormalizable. However, a perturbative calculation
with this propagator could give a unitary S-matrix, which is defined
between physical particle asymptotic states, as shown in ref.[2,3].
In the case of $d=4-\epsilon$, we can also choose the integration
contour of the Wick-rotated internal momenta so that it is consistent
with the Lee-Wick prescription. So we can get a unitary S-matrix in
our model, but it has non-analytic singularities which leads
to an acausality in a negligibly small-scale (proportional to the inverse of
the unphysical complex mass). Furthermore, any non-perturbative
calculational method, which implements the Lee-Wick prescription, is
not known. Then it is obscure that the 1/N expansion scheme
with Lee-Wick prescription could give a consistent perturbation
up to the higher orders. This is here an open problem.\par
  Our purpose
is here not to resolve these problems but to investigate the quantum
gravitational aspect from the following viewpoint.
Since the original lagrangian in $d=4-\epsilon$ does not contain the higher
derivative terms, then unphysical poles do not exist at all. But we
consider here a special resummation (1/N-expansion)
of the perturbation to get a convergent
result. Lee-Wick prescription was needed to justify this resummation.
Even if this prescription leads to an inconsistency
of the expansion at higher orders, we can see the aspect of quantum
gravity at least in the lowest (one loop) order
without breaking the unitarity
as far as we consider in $d=4-\epsilon$. \par
 As for the fractal dimension, there is an interesting idea [5] that
the dimension, $d=4-\epsilon$, would be effectively realized due to
the non-perturbative effect of the virtual and real black-holes. But
we do not investigate on this point here, and we consider the world of $d=4$
being realistic.
\par
 It should be noticed that we do not need to introduce any counter terms
in the step (ii) of taking the four-dimensional limit for the calculated
quantities, since there appears no divergence like $1/\epsilon$ in our
scheme. In other words, any
subtraction is not necessary contrary to the the case of the usual dimensional
regularization.
The reason why any divergence does not occur in the limit of $\epsilon=0$ is
in the fact that the expansion parameter in our scheme
is $\epsilon /N$ (not merely $1/N$).
As a result, the dangerous term like (finite term)$/\epsilon$, which
diverges in the usual dimensional regularization, becomes finite since the
expansion parameter cancels this singular factor $1/\epsilon$.
But the terms, which should remain in the usual renormalization procedure,
vanish in our scheme.
Thus we obtain an unusual result, but
we need not to introduce curvature squared terms as the counter terms even at
the four dimensional limit.
 This situation can be also understood from the four-dimensional limit of our
effective lagrangian.
Since any counter term is not introduced in taking the limit, the terms
proportional to $1/\epsilon$ are remained in the effective lagrangian.
If there is a term (say A) with the infinite coefficient in the action,
the constraint, A=0, will be obtained from the viewpoint of the path-integral.
In our formalism, we will therefore get a constraint, (the terms proportional
to $1/\epsilon$) = 0, in the limit of $\epsilon=0$. For example, we have
$\int d^4x\sqrt{-g}(R_{\mu\nu}R^{\mu\nu}-R^2/3)=0$. So it can be said that to
take the four dimensional limit of the quantities calculated in our scheme at
$d=4-\epsilon $
is equivalent to performing an usual 1/N-expansion at $d=4$ with the
constraint by which the higher derivative terms are excluded.
But the latter calculations would be a formidable task.
 We can see another calculational method [6] with a special regularization
procedure, which is applicable to a non-renormalizable theory, e.g. the
gravitational theory.
In this approach [6,7], the terms which are given in the usual subtraction
scheme are obtained as a final result. But such terms vanish in our approach
as stated above. This disagreement with our results comes from the difference
of the lagrangian used in the perturbation and the expansion parameter. \par
 Recently, an $\epsilon$-expansion has been performed near two dimension [8],
and it has been shown that
the Einstein theory has a non-trivial fixed point if there are sufficiently
many matter fields, and the physical quantity calculated in $d=2+\epsilon$ is
continued to $d=2$ by taking the limit of $\epsilon\to 0$. And it is shown [8]
that the anomalous dimensions of operators coincide with that of the Liouville
theory, which can be obtained as the trace anomaly in $d=2$ gravity.
It should be noticed that the calculation in $d=2+\epsilon$ is performed by
the metric fluctuations being absent at $d=2$. Nevertheless, the $\epsilon=0$
limit of the calculated quantity could provide a correct result obtained in
the Liouville theory. This is one of our motivations to study the physical
aspects of the four-dimensional gravity from $4-\epsilon$ dimension. \par
 Here we study the quantum-gravitational effects especially on the effective
potential of a scalar-field ($\phi$) according to the procedure stated above.
The possibility of the dynamical mass-generation caused by the quantum effect
of gravity is examined.
And we conclude that a mass-scale, the non-zero expectation value $<\phi>$,
can not be generated by the quantum gravitational effect both
at $d=4-\epsilon$ and  at the limit of $d=4$.
 \par
 As stated above, it has been found that the quantum gravitational-effect on
the field-operators of mass-dimension four (for example $\phi^4$) remains
finite but the term like $\phi^4\ln(\phi^2)$ vanishes in the limit of
four-dimension. Then the curvature-squared terms, $R_{\mu\nu}R^{\mu\nu}$ and
$R^2$, are induced as a quantum effect, but they should not be used in the
perturbation because they are not the fundamental terms of the original
lagrangian. In the next section, we study $\phi^4$-theory in $d=4-\epsilon$ to
apply our calculational method
and to examine its limit of $d=4$. In section {\bf 3}, the terms induced from
the matter loop-corrections are calculated in a coordinate
invariant way and the existence of a fixed point is assured for the effective
theory within the 1/N-expansion. In section {\bf 4}, we show that
the effective theory ($L_{eff}^{3/2}$), which includes the loop-corrections of
N spin-3/2 fields, should be used in the perturbation at $d=4-\epsilon$.
And the effective potential of a scalar-field is examined in section
{\bf 5} in terms of this effective lagrangian. The conclusion is given in the
last section. \par
\smallskip
\bigskip
\noindent{\bf 2. Renormalization in d=4-$\epsilon$}
\bigskip
 For the theory of $d=4-\epsilon$, we consider the following renormalization
procedure [4] even if the theory is not renormalizable: (a)
Introduce an ultraviolet cutoff, $\Lambda$. (b) Scale the dimensionful
parameters of the theory by the appropriate powers of $\Lambda$ times bare
dimensionless parameters. (c) Compute the appropriate quantities to leading
order with respect to the perturbative expansion parameter. (d) Require
conditions of the bare dimensionless parameters such that the theory is finite
and independent of $\Lambda$ as $\Lambda\to \infty$.\par
The conditions in the step (d) determines the critical trajectories,
which go to a non-trivial fixed point as $\Lambda\to\infty$, of dimensionless
bare-parameters defined in the step (b). Although our purpose is to apply the
above procedure to the gravitational theory, we firstly consider this method
in the simple $\phi^4$-theory in order to investigate the characteristic
properties of $4-\epsilon$ dimensional
theory. Consider the following lagrangian,
$$L_{\phi}={1 \over 2}(\partial_{\mu}\phi\partial^{\mu}\phi-m^2\phi^2)
            -{1 \over 12}\lambda\phi^4, \tag$1$ $$
\noindent where $m$ and $\lambda$ are the mass and the coupling constant,
then apply the above procedure by calculating the effective potential to
leading order in $\lambda$. \par
 Since we are interested in the dynamical mass-generation, we calculate the
effective potential of $\phi$ by assuming that $m=0$. According to the proper
time formulation, we get the following one-loop correction ($V^{(1)}$) to the
potential,
$$V^{(1)}={i \over 2}\mathop{lim}_{\eta \to 0}\int_{\eta}^{\infty}{ds \over s}
         \int {d^dk \over (2\pi)^d}
         exp(-iHs),  \tag$2$ $$

\noindent where
$$H=k^2+\lambda\phi^2$$
and $\eta(=\Lambda^{-2})$ is the ultraviolet cutoff-parameter. After
rotating the integral contour, we obtain
$$\align V^{(1)}&=-(\lambda\phi^2)^d{\pi^{d/2} \over 2(2\pi)^d}
                   \mathop{lim}_{\eta \to 0}\int_{\eta\phi^2}^{\infty}
                   {dx \over x}x^{-d/2}e^{-x} \tag$3$ \\
          &=-{\pi^{d/2} \over 2(2\pi)^d}\mathop{lim}_{\eta \to 0}
            \biggl\{\biggl[ {1 \over 2-\epsilon/2}\eta^{-2+\epsilon /2}
            -{(\lambda\phi^2) \over 1-\epsilon/2}\eta^{-1+\epsilon /2}
            -{(\lambda\phi^2)^2 \over \epsilon/2}\eta^{\epsilon /2}\biggr] \\
          &\qquad\qquad\qquad +(\lambda\phi^2)^{d/2}\Gamma (-{d \over 2})
           + O(\eta) \biggr\}. \tag$4$
\endalign$$
\bigskip
 The first two terms in Eq.(4) are divergent as $\eta\to 0$. But the
third term vanishes as $\eta\to 0$ for positive $\epsilon$, so the
$\phi^4$-term is not affected by the renormalization.
In order to get a conditions for $\Lambda\to\infty$ according to the
procedure stated above, each parameters are rescaled as follows,

$$ \lambda=\lambda_0\Lambda^{-\epsilon}, \quad m=m_0\Lambda^2,
                        \quad \eta=\Lambda^{-2}, \tag$5$ $$
where $\lambda_0$ and $m_0$ are dimensionless and we assumed that $m_0=0$.
Then we require
$$
   \biggl(m_0^2-{1 \over 16\pi^2}{\lambda_0 \over 1+\epsilon/2}
   \biggr)\Lambda^2=m_R^2\mu^2,  \tag$6$ $$

\noindent where $m_R$ is a renormalized, $\Lambda$-independent mass and
$\mu$ is a finite mass-scale.
Since we assumed $m_0=0$, we can see from Eq.(6) that the fixed point of
$\lambda_0$ (denoted by $\lambda^{\ast}$) is zero,

$$ \lambda^{\ast}=0  \tag$7$ $$

\noindent It is known [9] that this theory has an ultraviolet fixed point at
$\lambda=0$ and an infrared fixed point at $\lambda=2\epsilon/3$ within
$\epsilon\lambda$ expansion. So Eq.(7) is a reasonable result, and we should
consider the range of $\lambda$ being between 0
and $2\epsilon/3$.
  Making a fine tuning as $m_R=0$, we finally obtain the following potential
with one-loop correction,

$$\align V_{eff}&={1 \over 12}\lambda\phi^4
           -{\pi^{d/2} \over 2(2\pi)^d}(\lambda\phi^2)^{d/2}
           \Gamma \biggl( -{d \over 2}\biggr) \tag$8$ \\
          &={1 \over 12}\lambda\phi^4+
{(\lambda\phi^2)^2 \over 32\pi^2}{1 \over \epsilon}+{1 \over 64\pi^2}
            (\lambda\phi^2)^2[\ln\phi^2+const.]+O(\epsilon). \tag$9$
\endalign$$
\bigskip
 We can see from Eq.(8) the property of $V_{eff}$ in $d=4-\epsilon$. Since
$d<4$, the second term ($\propto\phi^d$) dominates over the first tree term
($\propto \phi^4$) near $\phi\sim 0$ no matter how small the value of
$\lambda$ is. Further $\Gamma(-d/2)>0$, so $V_{eff}$ has a non-trivial minimum
at $\phi=\phi_0$, which is given as follows,
$$\phi_0^{-\epsilon}\lambda^{1-\epsilon/2}={8\pi^2 \over 3\Gamma(-d/2)}
                              \tag$10$ $$
 \par
Next, we consider the limit of $\epsilon\to 0$, where $V_{eff}$ can be
expanded as Eq.(9), where the first three terms are remaining and the second
term seems being divergent. If we consider the usual renormalization scheme,
this divergent term should be subtracted. And we would obtain the following
renormalized effective potential ($V_{eff}^R$),

$$ V_{eff}^R={1 \over 12}\lambda\phi^4+
             {1 \over 64\pi^2}(\lambda\phi^2)^2
             [\ln\phi^2+c], $$
\bigskip
\noindent where $c$ is an appropriate constant.
However this is not the correct limit of $\epsilon=0$. Since
$0<\lambda<2\epsilon/3$
the value of $\lambda$ should be proportional to $\epsilon$, then $V_{eff}$
vanishes at the limit of $\epsilon=0$. On the other hand,
the minimum of Eq.(8), $\phi_0$ given in Eq.(10), can be approximated as
$\phi_0\propto 1/\sqrt{\epsilon}$. Then it moves to infinity in the limit of
$\epsilon=0$.
This is consistent with the fact that the theory approaches to a trivial
limit, the free theory. It is believed by many people that the four
dimensional $\phi^4$ theory is trivial, so the calculation at $d=4-\epsilon$
given here would be smoothly connected to the realistic limit of $d=4$.
 We will find a similar situation also in the gravitational theory as shown
below. \par

\bigskip
\noindent {\bf 3. Gravitational Model and Fixed Point}\par
\bigskip
Here, we consider the $1/N$ expansion scheme for the gravitation
to see the existence of a non-trivial ultraviolet fixed point and to
investigate the property of the induced term.
First, we consider the model, which contains a large number of spin-1/2 fields
($\Psi_i$, $i=1\sim N$), in $4-\epsilon$ dimension, as a prototype. \par
The lagrangian is given as,

$$L_{1/2}=\sqrt{-g}(\kappa^2R+\sum_{i=1}^N \bar\psi_i\gamma^{\mu}D_{\mu}
          \psi_i+\lambda). \tag$11$ $$
$$D_{\mu}=\partial_{\mu}+{1 \over 2}\omega_{\mu}^{ab}\sigma^{ab},
          \qquad\quad \sigma^{ab}={1 \over 4}[\gamma^a,\gamma^b]  $$
\smallskip

\noindent From Eq.(11), the effective lagrangian ($L_{eff}^{1/2}$, the
Eq.(31)) can be obtained by integrating N fermions.
If this model were considered in four-dimension, the higher derivative terms,
$R_{\mu\nu}R^{\mu\nu}$ and $R^2$, must be added to $L_{1/2}$ in order to make
the theory renormalizable. However, the terms proportional to $1/\epsilon$
are not singular at $d=4-\epsilon$, so unfavorable counter terms like
$R_{\mu\nu}R^{\mu\nu}$ and $R^2$ are not needed here. Then the theory given
by Eq.(11) in $d=4-\epsilon$ is free from the problem of the ghost.
Furthermore, this theory is renormalizable if the perturbation were performed
within the 1/N expansion scheme [5], and a non-trivial fixed point exists.
However $L_{eff}^{1/2}$ is not available in the perturbation because of the
existence of the tachyonic pole in the propagator as shown below.  \par
 In the previous calculation [4] of the fermion-loop,
the coordinate non-invariant regularization was used.
Then the effective lagrangian , $L_{eff}^{1/2}$, obtained in this way has lost
the general-coordinate invariance. This situation complicates matters when we
try to derive a physical consequence from $L_{eff}^{1/2}$, so we recalculate
here the fermion loop according to a gauge invariant regularization, namely
the Pauli-Villars method. \par
 By expanding the metric as $g_{\mu\nu}=\eta_{\mu\nu}+h_{\mu\nu}$,
we calculate the self-energy part of $h_{\mu\nu}$ due to the fermion-loop.
Let the s-th regulator's mass as $m_s$.
After summing up the loops of a massless fermion and a number of massive
regulators, the resultant self-energy $\Pi^{(1)}_{\mu\nu\lambda\sigma}$ can
be written as

$$ \Pi_{\mu\nu\lambda\sigma}^{(1)}=\sum_{s=0}\sum_{i=a,b}
           C_s\Pi_{\mu\nu\lambda\sigma}^{(i)}(p,m_s) \tag$12$ $$
\bigskip
\noindent where $C_0=1$ and $m_0=0$. The detail of the calculation
is given in the {\bf Appendix}, where we can see that the following conditions,

$$\sum_{s=0}C_s=0,\qquad \sum_{s=0}C_sm^2_s=0 \tag$13,14$ $$
\bigskip
\noindent are sufficient to eliminate the divergences in $\Pi^{(a,b)}$.
Using these, the final result is obtained as

$$ \Pi_{\mu\nu\lambda\sigma}^{(1)}
           ={{\Gamma(2-d/2)} \over {(4\pi)^{d/2}}}
            {tr1 \over 2}\biggl\{
            {(p^2)^{-\epsilon/2} \over 4}
            B({{d-2} \over 2},{{d-2} \over 2})
            {\tilde \Pi}_{\mu\nu\lambda\sigma}^{(1,0)}
                     -\sum_{s=1}{(m_s^2)^{-\epsilon/2} \over d-2}C_s
           {\tilde \Pi}_{\mu\nu\lambda\sigma}^{(1,m_s)}
            \biggr\} \tag$15$ $$
\bigskip
\noindent where tr1 denotes the dimension of spinor-space and

$$\align &{\tilde \Pi}_{1/2}^{(0)}\equiv
        h^{\mu\nu}{\tilde \Pi}_{\mu\nu\lambda\sigma}^{(1,0)}h^{\lambda\sigma}
           ={1 \over d+1}W^{(2)}
           +{d-4 \over 12(d^2-1)}\biggl(\sqrt{g}R^2\biggr)^{(2)},  \tag$16$ \\
         &{\tilde \Pi}_{1/2}^{(m)}\equiv
          h^{\mu\nu}{\tilde \Pi}_{\mu\nu\lambda\sigma}^{(1,m)}h^{\lambda\sigma}
           ={{m^2} \over 3}\biggl(\sqrt{g}R\biggr)^{(2)}
           +{{8m^4} \over d}\biggl(\sqrt{g}\biggr)^{(2)}+O(m^0). \tag$17$
\endalign$$
\bigskip
\noindent where

$$  W=R_{\mu\nu}R^{\mu\nu}-{1 \over d-1}R^2.  $$

\bigskip
\noindent Here, $(\sqrt{g}R)^{(2)}$ e.t.c. with the upper subscript ${}^{(2)}$
represent the quadratic part of $\sqrt{g}R$ e.t.c. with respect to
$h_{\mu\nu}$ in the momentum representation. $\tilde\Pi^{(1,m)}$ and
$\tilde\Pi^{(1,0)}$ denote the self-energy part coming from the massive
regulators and the massless fermion of $\Pi^{(1)}$ respectively. \par
 Eq.(17) implies that only the two parameters, $\kappa$ and $\lambda$ in
$L_{1/2}$, are renormalized when each regulator mass, $m_s$, are
taken to be infinite. And we need not any other counter term.
As in the previous section, we nextly replace the parameters as

$$ \kappa^2=\kappa^2_0\Lambda^{2-\epsilon},\qquad
             \lambda=\lambda_0\Lambda^{4-\epsilon}, \qquad
             m_s=\xi_s\Lambda. \tag$18,19,20$ $$

\noindent Then, we adjust them by requiring

$$ \biggl(\kappa_0^2-{{\Gamma(2-d/2)} \over {(4\pi)^{d/2}}}
         {tr1 \over 6(d-2)}\sum_{s=1}C_s\xi_s^{2-\epsilon}
   \biggr)\Lambda^{2-\epsilon}=\kappa^2_R\mu^{2-\epsilon},
   \tag$21$ $$

$$ \biggl(\lambda_0-{{\Gamma(2-d/2)} \over {(4\pi)^{d/2}}}
         {4tr1 \over d(d-2)}\sum_{s=1}C_s\xi_s^{4-\epsilon}
   \biggr)\Lambda^{4-\epsilon}=\lambda_R\mu^{4-\epsilon},
   \tag$22$ $$
\bigskip
\noindent where $\kappa_R$ and $\lambda_R$ are finite, $\Lambda$ independent
numbers. These equations imply the following fixed point for
each dimensionless parameter,

$$ \kappa^{*2}={{\Gamma(2-d/2)} \over {(4\pi)^{d/2}}}
         {tr1 \over 6(d-2)}\sum_{s=1}C_s\xi_s^{2-\epsilon}, \qquad
  \lambda^*={{\Gamma(2-d/2)} \over {(4\pi)^{d/2}}}
         {4tr1 \over d(d-2)}\sum_{s=1}C_s\xi_s^{4-\epsilon}.
   \tag$23,24$  $$

\noindent Both $\kappa^*$ and $\lambda^*$ seem to be divergent at $d=4$ due
to the factor $\Gamma(2-d/2)$, but it can be seen that their values are
finite if we use Eq.(14) and the extra requirement,
$$\sum_{s=1}C_s\xi_s^4=0, \tag$25$ $$
is imposed. Then we obtain the following results in d=4 limit,

$$   \kappa^{*2}=-{tr1 \over 96\pi^2}
                \sum_{s=1}C_s\xi_s^2\ln\xi_s, \qquad
        \lambda^*=-{tr1 \over 16\pi^2}
                \sum_{s=1}C_s\xi_s^4\ln\xi_s. \tag$26,27$ $$
\bigskip
\noindent These results are qualitatively consistent with that given
in [4], where a gauge non-invariant cutoff was used.
The absolute values of $\kappa^*$ and $\lambda^*$ can be set appropriately
, since they depend on the regularization scheme. Our purpose is here to
assure the existence of a non-trivial fixed point and not to examine their
values, we do not discuss on this point further more. In any case, the first
two terms given in Eq.(17) are regularized by the parameters in $L_{1/2}$,
and the terms given in Eq.(16) remain as the
induced terms. In the next section, we discuss on these induced terms, which
play an important role in the effective lagrangian. \par
 For the sake of the later convenience, we give the loop corrections
of spin-3/2 field, which is denoted by $\Psi_{\mu}(x)$.
For this field, the loop correction is written as

$$ Tr[\ln(\gamma^{\mu}D_{\mu})_{3/2}-3\ln(\gamma^{\mu}D_{\mu})_{1/2}],
                                       \tag$28$ $$
\bigskip
\noindent where the lower suffix 3/2 (1/2) means that $\gamma^{\mu}D_{\mu}$
operates on spin-3/2 (-1/2) field. Eq.(28) is obtained [10] by imposing the
gauge condition, $\gamma^{\mu}\Psi_{\mu}=0$, and the
constraint, $D^{\mu}\Psi_{\mu}=0$. These condition and constraint provide the
second term of Eq.(28).
For Eq.(29), we calculate $\Pi_{\mu\nu\lambda\sigma}^{(1)}$ by introducing
massive regulators as in the case of the loop-correction of spin-1/2 field.
Imposing the conditions Eqs.(13,14), we obtain the results which are written
in the form of Eq.(15) where ${\tilde \Pi}_{\mu\nu\lambda\sigma}^{(1,0)}$ and
${\tilde \Pi}_{\mu\nu\lambda\sigma}^{(1,m)}$ are changed as,

$$\align {\tilde \Pi}_{3/2}^{(0)}\equiv
        h^{\mu\nu}{\tilde \Pi}_{\mu\nu\lambda\sigma}^{(1,0)}h^{\lambda\sigma}
          &=\biggl({d-3 \over d+1}-8{d-2 \over d-1}\biggr)W^{(2)} \\
          & +{1 \over 12(d-1)}\biggl[{(d-3)(d-4) \over d+1}-8(d-2)\biggr]
            \biggl(\sqrt{g}R^2\biggr)^{(2)},  \tag$29$ \\
         {\tilde \Pi}_{3/2}^{(m)}\equiv
          h^{\mu\nu}{\tilde \Pi}_{\mu\nu\lambda\sigma}^{(1,m)}h^{\lambda\sigma}
          &={d-3 \over 3}m^2\biggl(\sqrt{g}R\biggr)^{(2)}
           -{24 \over d}m^4\biggl(\sqrt{g}\biggr)^{(2)}+O(m^0). \tag$30$
\endalign$$
\bigskip
 Here we comment on the results of Eqs.(29) and (30); (i) In Eq.(29),
the coefficient of $W^{(2)}$ is negative near $d=4$ contrary to Eq.(16) of
spin-1/2. (ii) The values of $\kappa^*$ and $\lambda^*$ in this case are
obtained by multiplying $d-3$ and -3 to the values given
in Eqs.(23) and (24) respectively. (iii) We may use the square of
$R_{\mu\nu\lambda\sigma}$ or the Weyl curvature, $C_{\mu\nu\lambda\sigma}$,
in Eq.(29) instead of $W$, or we can use both. But it is not essential here
how we write Eq.(29), so we use the above expression. \par
 It should be noticed that
the first point is the essential difference between the cases of spin-1/2 and
spin-3/2 fields if we consider the 1/N-expansion scheme. On
this point, we will discuss in the next section. \par
\bigskip
\noindent{\bf 4. Effective Action in 4-$\epsilon$ Dimensional Gravity}
\bigskip
Performing the integration of N spin-s fields and the renormalization of
$\kappa$ and $\lambda$ as shown above, we obtain the effective lagrangian,
$L_{eff}^{s}$. Firstly, we consider the case of $s=1/2$ to find what kind of
effective lagrangian is suitable to the perturbation. The effective
lagrangian is written as follows,

$$L_{eff}^{1/2}=N\sqrt{-g}\biggl(\kappa^2R+\lambda+Tr\ln_R(O_{1/2})
              \biggr), \tag$31$ $$
\bigskip
\noindent where $\kappa=\kappa_R\mu^{1-\epsilon}/N^{1/2}$,
$\lambda=\lambda_R\mu^{4-\epsilon}/N$ and $O_{1/2}=\gamma^\mu D_\mu$.
$Tr\ln_R(O_{1/2})$ denotes the regularized fermion-loop correction.
{}From the result obtained in section {\bf 3}, Eqs.(15) and (16), the third
term
can be written as,
$$ A_{1/2}(d)(p^2)^{-\epsilon/2}
          \biggl(W^{(2)}+B_{1/2}(d)(R^2)^{(2)}\biggr), \tag$32$ $$
\noindent where $B_{1/2}(d)=(d-4)/[12(d-1)]$ and
$W$ is the Weyl term in the d-dimension defined in the previous section.
The proportional factor $A_{1/2}(d)$ is written as,

$$     A_{1/2}(d)\equiv\Gamma(2-d/2)\beta_{1/2}(d), \tag$33$ $$
\bigskip
\noindent where

$$     \beta_{1/2}(d)={1 \over {(4\pi)^{d/2}}}
            {tr1 \over 8(d+1)}
             B({{d-2} \over 2},{{d-2} \over 2}). \tag$34$ $$
\bigskip
\noindent Here we should notice that $A_{1/2}(d)$ is divergent at $d=4$ due
to the factor $\Gamma(2-d/2)$ since $\beta_{1/2}(4)(=1/160\pi^2)$ is finite
at this limit. \par
 The above results, Eqs.(32$\sim$ 34), can be generalized to the case of
other matter-field of spin-s with a similar formula by replacing
the suffix ${}_{1/2}$ of $B_{1/2}(d),\beta_{1/2}(d),A_{1/2}(d)$ and
$L_{eff}^{1/2}$ by s. Of course, those functions of d depend on the spin of
the fields. For the case of $s=3/2$, we obtain from Eq.(29),

$$     \beta_{3/2}(d)={1 \over {(4\pi)^{d/2}}}
            {tr1 \over 8}\biggl({d-3 \over d+1}-8{d-2 \over d-1}\biggr)
             B({{d-2} \over 2},{{d-2} \over 2}). \tag$35$ $$
\bigskip
\noindent From this, the four dimensional limit $\beta(4)$ can be written as,

$$ \beta_{3/2}(4)=-{77 \over 3}\beta_{1/2}(4).  \tag$36$ $$
\bigskip
\noindent Here we should notice that the sign of $\beta_{3/2}(4)$ is
opposite to the one of spin-1/2. This point is important as seen below. \par
 Nextly, we investigate the propagator of $h_{\mu\nu}$ derived from
$L_{eff}^s$. The spin-two part of the propagator, which is denoted as
$D_{\mu\nu\lambda\sigma}(p)$ with the momentum $p$, is given as follows,

$$ D_{\mu\nu\lambda\sigma}(p)={P^{(2)}_{\mu\nu\lambda\sigma}(p) \over Np^2
                \biggl(\kappa^2-A_s(d)(-p^2)^{1-\epsilon/2}
                              \biggr)}\,\, \tag$37$ $$
\noindent where the momentum is rotated to the minkowskian and
$P^{(2)}_{\mu\nu\lambda\sigma}(p)$ represents the spin-two projection
operator in d-dimension, which is given as
$$\align P^{(2)}_{\mu\nu\lambda\sigma}(p)&
               \theta_{\mu\lambda}\theta_{\nu\sigma}
              -{1 \over n-1}\theta_{\mu\nu}\theta_{\lambda\sigma}, \tag$38$ \\
        \theta_{\mu\nu}&=\eta{\mu\nu}-{p_{\mu}p_{\nu} \over p^2}.
\endalign $$
\bigskip
\noindent Here we assumed that $\lambda=0$. As seen in the followings, the
property of $D_{\mu\nu\lambda\sigma}$ is sensitive to the sign of $A_s(d)$.
\par
 For a while, we consider the four dimensional limit of Eq.(37). If we take
the limit of $\epsilon=0$, the denominator of Eq.(37) is written as

$$ p^2\biggl(\kappa^2-\beta_s(4)p^2[ln(-p^2)+{2 \over \epsilon}]\biggr) .
                                         \tag$39$ $$
\bigskip

\noindent In this case, we must add ${1 \over \alpha}W$, where $\alpha$ is
the parameter to be renormalized, to $L_{eff}^s$ in order to subtract the
divergent term (proportional to $1/\epsilon$) in Eq.(39). Then the theory is
changed to the usual higher derivative theory. In this case, the renormalized
form of Eq.(39) can be written as,
$$ p^2\biggl(\kappa^2-\beta_s(4)p^2\ln(-p^2/\mu^2)\biggr),
                                         \tag$40$ $$
\noindent where $\mu$ is an appropriate constant. For the case of
$\beta_s(4)>0$, the propagator has no real pole except for $p^2=0$ in the
eucledian, and it possesses a pair of complex conjugate pole in the
minkowskian [2]. This is realized in the case of $s=1/2$, where
$\beta_{1/2}(4)>0$. Then the problem of unitarity can be evaded by the
Lee-Wick mechanism. So the N matter-fields should be chosen at d=4 so that
$\beta_s (4)$ is positive.\par
 On the contrary, the propagator in $4-\epsilon$ dimension (Eq.(37))
has a tachyonic pole if $A_s(4-\epsilon)$ is positive, the case of spin-1/2
field. Then we must select another appropriate matter-field which could
provide negative $A_s(4-\epsilon)$ in order to improve this difficulty.
The only candidate for such a field is the spin-3/2 field
(Rarita-Schwinger field), $\Psi_{\mu}$. And we can assure
$A_{3/2}(4-\epsilon)<0$ from eqs.(35) and (36). On the other hand, this means
that we could not use $L_{eff}^{3/2}$ at $d=4$. \par
 After all, the propagator obtained from $L_{eff}^{3/2}$ does not have a
tachyonic pole in $d=4-\epsilon$. And the pole other than $p^2=0$ is a pair
of complex conjugate pole in the physical plane, $p^2=re^{\pm \theta}$,
where

$$ r\simeq (\kappa^2/A_{3/2})^{1+\epsilon/2},\qquad \theta\simeq \pi\epsilon/2.
                    \tag$41,42$ $$
\bigskip
\noindent As mentioned in the introduction, we can perform a perturbation
keeping the unitarity in this case, at least at the lowest order.
 \par
The situation with respect to the regularization and the fixed
point does not change qualitatively in this case compared to the
case of spin-1/2 field. Then we use $L_{eff}^{3/2}$ in the following
perturbative calculation of the quantum gravitational effect in
$d=4-\epsilon$. In the next section, we use this model to investigate
the effective scalar-potential with one-loop correction of gravitational
fluctuation. \par
\bigskip
\noindent{\bf 5. Effective Potential and Spontaneous Mass Generation}
\bigskip
 Several authors [11] have studied whether the radiative correction of
gravitational modes could lead to a spontaneous symmetry breaking which has
been given by the radiative correction of a gauge field [12]. But all these
approaches are performed just on the four-dimension,
where no reliable perturbative theory of quantum gravity is not known. Here
we investigate the same problem at $d=4-\epsilon$ in terms of
$L_{eff}^{3/2}$ given above and the result is continued to $d=4$. For
convenience, the cosmological constant $\lambda$ is fine tuned to
be zero.\par
 We add to $L_{eff}^{3/2}$ the following scalar ($\phi$) part

$$ L_{\phi}=\sqrt{-g}\biggl({1 \over 2}g^{\mu\nu}\partial_{\mu}\phi
            \partial_{\nu}\phi+\zeta\phi^2R+N^{-1}\gamma \phi^4\biggr).
               \tag$43$ $$
\bigskip
\noindent For the later convenience, we scale $\phi$ as $\sqrt{N}\phi$ and
rewrite $L_{\phi}$ as,

$$ L_{\phi}=N\sqrt{-g}\biggl({1 \over 2}g^{\mu\nu}\partial_{\mu}\phi
            \partial_{\nu}\phi+\zeta\phi^2R+\gamma \phi^4\biggr). \tag$44$ $$
\bigskip
\noindent The gravitational loop-correction to the scalar-potential is
obtained by the formula Eq.(2) with a different quadratic-operator $H$
derived from the lagrangian,

$$ L_{eff}=L_{eff}^{3/2}+L_{\phi}.  \tag$45$ $$
\smallskip
\noindent In order to get a complete result, we must further add the
gauge fixing term of the general coordinate transformation and the ghost part
to $L_{eff}$. But the ghost part does not contribute to the
potential of $\phi$, and we can always make the gauge fixing term which does
not affect the spin-two part of the graviton propagator. So,
it is resonable to concentrate our attention on the spin-two component of
$h_{\mu\nu}$, which is gauge-independent, because the physical result should
be independent of the gauge condition. From Eq.(45), $H$ for spin-2 component
is given as follows,

$$ H=Gk^2+\alpha(d)(k^2)^{d/2}+\gamma\phi^4 \tag$46$ $$
\smallskip
\noindent where $\alpha(d)\equiv-A(d)_{3/2}(>0)$, and

$$ G=\kappa^2+\zeta\phi^2 \tag$47$ $$
\bigskip
{\bf (5-1)} Before investigating the general case of $\alpha(d)\neq 0$, we
firstly examine the case of no induced term, e.g. $\alpha(d)=0$.
This is corresponding to the usual perturbation in the 4-$\epsilon$
dimensional Einstein gravity. In this case, the theory is not renormalizable
and
it would be difficult to find a non-trivial fixed point. \par
 In this case,
the gravitational one-loop correction to the potential can be calculated by
taking N=1 in Eq.(43) as,

$$ V^{(1)}=-P(d)\Gamma({d \over 2})
           \Gamma(-{d \over 2})
           ({\gamma \over G}\phi^4)^{{d \over 2}}, \tag$48$ $$
\bigskip
\noindent where

$$ P(d)={\pi^{d/2} \over 2(2\pi)^d\Gamma(d/2)}. \tag$49$ $$
\bigskip
\noindent We study this potential in the following two cases; (a) The
scalar field couples minimally to gravity, $\zeta=0$ and $G=\kappa^2$. (b)
$\zeta\neq 0$ but $\kappa=0$, where the Einstein term should be produced by
the condensation of the scalar field. \par
The purpose of considering the case of (a) is to investigate the possibility
of the dynamical symmetry breaking due to the gravitational loop-correction.
There, Eq.(48) is written as

$$ V^{(1)}=-P(d)\Gamma({d \over 2})
           \Gamma(-{d \over 2})
           ({\gamma \over \kappa^2}\phi^4)^{{d \over 2}} \tag$50$ $$
\bigskip
\noindent This potential decreases with $\phi$. However, since it is
proportional to $(\phi^4)^{2-\epsilon/2}$, the tree part,
$V^{(0)}=\gamma\phi^4$, dominates over $V^{(1)}$ at small $\phi$. So it could
not cause any symmetry breaking since it gives a trivial solution,
$<\phi>=0$. Even if $\zeta\neq 0$, we obtain the same result with Eq.(43a) at
sufficiently small $\phi$, since $G=\kappa^2+\zeta\phi^2\sim \kappa^2$. \par
Next, we consider the case (b), where $\kappa$ is set to be zero, to
investigate whether the gravitational constant $\kappa$ could be generated by
non-zero $<\phi>$. In this case, $G=\zeta\phi^2 $ and we get the following
potential,

$$ V^{(1)}=-P(d)\Gamma({d \over 2})
           \Gamma(-{d \over 2})
           ({\gamma \over \zeta}\phi^2)^{{d \over 2}} \tag$51$ $$
\bigskip
\noindent Since the power of $\phi$ in $V^{(1)}$ is smaller than four
contrary to the case of (a), this potential dominates the tree potential
($V^{(0)}=\zeta\phi^4$) near $\phi\sim 0$. And $V^{(1)}$ is negative, so we
have a nontrivial vacuum, namely $<\phi>\neq 0$. This result is corresponding
to the case given in section {\bf 2}, Eq.(8). However, since this calculation
is based on the perturbative expansion with respect to $(\zeta<\phi>^2)^{-1}$
and the graviton's propagator behaves as $1/p^2$, it is difficult to find a
reliable ultraviolet fixed point because of the non-renormalizablity.
In other words, we can not assure the continuous limit of the
theory in this calculation.So we could not derive any definite
conclusion from this one-loop approximation. \par
 {\bf (5-2)} Now we turn to the case of $\alpha(d)\neq 0$, the 1/N-expansion
scheme. For $\alpha(d)\neq 0$, the integration of Eq.(2) can
not be written in a simple form.
So we represent the result of the integration by the sum of the series of
$1/\alpha(d)$,
since $\alpha(d)$ is very large near $d\sim 4$.
After a simple calculation, we obtain

$$ V^{(1)}=-P'(d){1 \over \alpha(d)}\sum_{n=0}^{\infty}{(-G)^n \over n!}
                 \alpha^{-{2 \over d}n}(d)\Gamma(1+{2 \over d}n)
                 I(n;\phi), \tag$52$ $$
\bigskip
\noindent where $P'(d)=2P(d)/d$ and

$$ I(n;\phi)=\mathop{lim}_{\eta \to 0}\int_{\eta}^{\infty}
                   {ds \over s}s^{(1-2/d)n-1}e^{-\gamma\phi^4s}. \tag$53$ $$
\smallskip
\noindent In this integration, the ultraviolet cutoff $\eta$ is introduced
since the integral diverges for $n=0,1$ and $2$. For large $\alpha(d)$, the
most dominant term is that of $n=0$, so we can approximate $V^{(1)}$ as
follows.

$$ \align V^{(1)}&\sim-{2 \over d}P(d){\gamma \over \alpha(d)}\phi^4
                   \mathop{lim}_{\eta \to 0}\int_{\eta\gamma\phi^4}^{\infty}
                   {dx \over x}x^{-1}e^{-x} \tag$54$ \\
                 &=-{2 \over d}P(d){\gamma \over \alpha(d)}
                   \biggl[ \phi^4(\ln\phi^4-b)+\mathop{lim}_{\eta \to 0}
                   ({\eta^{-1} \over \gamma}+\phi^4\ln\eta)\biggr].  \tag$55$
\endalign$$
\bigskip
\noindent The terms which diverge at $\eta=0$ can be absorbed into the
cosmological constant and $\phi^4$-term by a suitable renormalization. The
remaining finite term in Eq.(55) has a similar form to that
of the Coleman-Weinberg potential induced by the radiative correction, but
its sign is opposite. Then the the minimum of the potential, $V=V^0+V^{(1)}$,
is the trivial one, $<\phi>=0$. This implies that the gravitational effect
can not bring on a spontaneous symmetry-breaking as in the gauge theories.
The result is similar to that of $\alpha(d)=0$ with Eq.(50), but the form
of the potential is fairy different. \par
 Next, we investigate the problem of spontaneous mass generation by setting
$\kappa=0$, then $G=\zeta\phi^2$. If $\gamma\neq 0$, we have the same result
with Eq.(55). So $\kappa$ can not be spontaneously generated. For the case of
$\gamma=0$, the problem would be complicated due to the infrared divergences
if we consider the formula Eqs.(52,53). So we factorize the operator
$H$ as follows,

$$ Tr\ln H=Tr\biggl(\ln[-\partial^2]+\ln[(-\partial^2)^{-1+d/2}+\phi'^2]
                     \biggr), \tag$56$ $$
\bigskip
\noindent where $\phi'^2=\zeta\phi^2/\alpha(d)$. The first term gives the
$\phi$-independent contribution, which is infrared-divergent. Then it is
subtracted and only the second term is estimated by Eq.(2)
with $H=(k^2)^{-1+d/2}+\phi'^2$. After a calculation, we obtain

$$ V^{(1)}=-{2 \over d-2}P(d)\Gamma({2 \over d-2})\Gamma(-{d \over d-2})
            (\phi'^2)^{d/(d-2)}. \tag$57$ $$
\bigskip
\noindent Since $d/(d-2)\sim 2+\epsilon/2$ for $d=4-\epsilon$,
$\Gamma(-d/(d-2))<0$ so $V^{(1)}>0$. Then $V(=V^{(1)}$ for $\gamma=0)$ can
not have a minimum other than the trivial one at $\phi=0$. \par
Then, the gravitational constant $\kappa$ can not be spontaneously
generated by the quantum-gravitational effect.
This conclusion is different from the implication of Eq.(51) which is
obtained by a simple perturbation of the Einstein gravity including
a nonminimal coupling with a scalar.
The occurrence of this difference would imply the unreliability of the
usual perturbation in the gravitational theory. \par
\bigskip
{\bf (5-3)} Finally, we consider the limit of $\epsilon=0$ of $V^{(1)}$.
First, consider the potential given in {\bf (5-1)}, the case of
$\alpha(d)=0$. In this limit, eqs.(50) and (51) can be written as follows,

$$ V^{(1)}={1 \over 64\pi^2}\biggl({\gamma \over \kappa^2}\biggr)^2
            \phi^8\biggl(\ln(\phi^2/\mu^2)-{1 \over \epsilon}
             +O(\epsilon)\biggr), \tag$50'$ $$
\smallskip

$$ V^{(1)}={1 \over 64\pi^2}\biggl({\gamma \over \zeta}\biggr)^2
            \phi^4\biggl(\ln(\phi^2/\mu^2)-{1 \over \epsilon}
             +O(\epsilon)\biggr), \tag$51'$ $$
\bigskip
\noindent where $\mu$ denotes an appropriate scale factor. The divergent term
proportional to $1/\epsilon$ should be subtracted
by the counter-terms, $\phi^8$ and $\phi^4$ for $(50')$ and $(51')$
respectively.
The $\phi^8$ term is needed in the case of $\zeta=0$ due to the one-loop
diagram shown in Fig.1(a). The loop-diagram of Fig.1(b) is included in Eq.(51).
\bigskip
\centerline {\bf Fig.\ 1(a), (b)}
\bigskip
\noindent After the subtraction, we obtain the finite term, the first term
in eqs.$(50')$ and $(51')$. The latter potential, Eq.$(51')$, is the
Coleman-Weinberg type, and we may expect the occurrence of a spontaneous mass
generation due to the loop-correction of gravitation. But we should notice
that this calculation is based on the non-renormalizable perturbation, by
which it is difficult to perform a renormalization group analysis. Then we
could not trust this perturbative result. \par
 Nextly, we consider Eq.(55), the case of $\alpha(d)\neq 0$. The limit of
the prefactor is written as follows,

$$ {2 \over d}P(d){\gamma \over \alpha(d)}
                  ={15\gamma \over 308}\epsilon +O(\epsilon^2). \tag$58$ $$
\bigskip
\noindent Since the remaining part of Eq.(55) is independent of $\epsilon$,
the potential $V^{(1)}$ seems to vanish in the limit of $\epsilon=0$ due to
this
prefactor. However we must be careful in taking the limits of $\epsilon=0$ and
$\eta=0$ for the third term in the parenthesis. The divergent term,
$\phi^4\ln\eta$, comes from the Fig.2(a).
\bigskip
\centerline {\bf Fig.\ 2(a), (b), (c)}
\bigskip
\noindent The loop-integral of the graviton leads to the logarithmic
divergence, since the integral can be written as $\int d^4k/k^4$ at large $k$
in the four-dimensional limit. If we use dimensional regularization, this
divergence is like $1/\epsilon$. Then $\epsilon\ln\eta$ should be finite for
$\epsilon$, $\eta\rightarrow 0$.
If we set
    $$\epsilon\ln\eta=-c_1, \tag$ 59$  $$
\bigskip
\noindent we obtain the following four dimensional limit,

$$\mathop{lim}_{\epsilon\to 0}V^{(1)}={15\gamma \over 308}c_1\phi^4. \tag$60$
$$
\bigskip
\noindent We should notice here that this result is obtained without
any
subtraction. But this gives rise to a finite modification of the tree
parameter $\gamma$ by the order of $O(1/N)$. \par
Then it can be said that the quantum effect of gravitation in four dimension
could not provide a
Coleman-Weinberg type potential, which gives rise to a spontaneous symmetry
breaking and generate some dynamical mass.
So the other effect like the radiative correction of a gauge field would be
necessary to give a non-trivial vacuum expectation value of a
scalar. \par
 Before considering another quantity, which is affected
by the quantum fluctuation of the gravity at $d=4$, we consider the
reason of obtaining the result of Eq.(60).
The calculation given here is the perturbation with respect to the parameter,
1/$\sqrt{N\alpha(d)}$, which is the inverse of the coefficient of the induced
term coming from the loop-correction. Choosing this expansion-parameter is
reasonable since the induced term is dominant in the ultraviolet region
and the parameter is very small. So the metric should be expanded as

$$ g_{\mu\nu}=\eta_{\mu\nu}+{1 \over \sqrt{N\alpha(d)}}h_{\mu\nu}, \tag$61$ $$
\bigskip
\noindent in the perturbation.
This is equivalent to investigate the zero-coupling limit of the four
dimensional theory, because $\sqrt{1/\alpha(d)}\propto \sqrt{\epsilon}$.
Namely, we are considering the $\epsilon/N$-expansion near $d=4$.
However in some diagram, this zero of the expansion parameter can be
cancelled by the ultraviolet divergence of $O(1/\epsilon)$ coming from the
integration of a loop-momentum as shown in the above example, Eq.(60). Then
only the diagram, which includes an ultraviolet divergence, could provide a
finite result in the limit of $\epsilon=0$. But
the term, which should remain finite in the usual calculation of the
Feynman diagram, vanishes. \par
  Here we should notice another important point how the higher derivative
terms are avoided in our calculation. The induced term in $L_{eff}^s$ can be
expanded as, $A/\epsilon+B+\epsilon C+O({\epsilon}^2)$,
and $A=\int d^4x \sqrt{-g}(R_{\mu\nu}R^{\mu\nu}-R^2/3)$. Then it can be said
that the dangerous term, which provides the massive ghost, is excluded in
our scheme by the constraint,
$\int d^4x \sqrt{-g}(R_{\mu\nu}R^{\mu\nu}-R^2/3)=0$, in the limit of
$\epsilon=0$. This could be understood from the viewpoint of the
path-integral formulation. It should be noticed that this is not equivalent to
taking the limit of zero coupling constant in the four-dimensional higher
derivative theory. In this case, the massive ghost contributes to all the
calculated quantities. But in our case, ghost does
not exist at all since the expanded terms B,C and others are added to
$A/\epsilon$ to form a desirable induced term.
Then our scheme is equivalent to the calculation at $d=4$ where only the
finite part of the induced term is included in $L_{eff}$ and the
constraint mentioned above is imposed at all stages of the perturbation.
This procedure seems to be very complicated compared to our approach from
$d=4-\epsilon$.
\par
 Next, we examine another example of the quantum-gravitational effect
which remains even at the limit of $d=4$. First,
consider the self energy of the graviton itself. The dominant Feynman
graph is shown in the Fig.2(b).
Here the vertex part with the matter-loop comes from the induced term,
which is proportional to $1/\sqrt{\alpha(d)}$. Because the three point
vertex coming from $\sqrt{g}R$ is proportional to $(1/\sqrt{\alpha(d)})^3$,
so it is suppressed here. The integration with respect to
the loop-momentum of this diagram gives rise to a divergence of the same
order with $1/\epsilon$ in the four dimensional limit. But
this divergence can be cancelled out due to the square of the effective
coupling constant, $1/\alpha(d)$. Then the corresponding counter-terms
which should be introduced to cancel the $1/\epsilon$ singularities are not
necessary. The finite terms obtained in this way contain
the higher derivative terms, $R_{\mu\nu}R^{\mu\nu}$ and $R^2$.
So we can say that these quadratic terms are generated by the quantum
effect of gravitation, because they are absent in the original lagrangian.
But the coefficients of them should be very small,
and they should not be used in the perturbation of the gravitation. This
is similar to the case of a renormalizable theory in the four dimension,
where we generally obtain many kinds of higher dimensional operators as
quantum effects.
\par
Due to the same reason as above, a similar finite contribution is obtained
for the self-energy diagram of gauge bosons. The corresponding
diagram is shown in the Fig.2(c).
In order to see the gauge independent part of this graph, we pick up
the vertex which survives when the legs of $h_{\mu\nu}$ and $A_{\mu}$ are
contracted by $P^{(2)}_{\mu\nu\lambda\sigma}$ and the transverse
operator, $\theta_{\mu\nu}=p_{\mu}p_{\nu}-\eta_{\mu\nu}/p^2$, respectively.
It is written as,

$$ -i{2 \over \sqrt{\alpha(d)}}\biggl[q_{\mu}q_{\nu}\delta_{\rho\sigma}
           -q_{\mu}p_{\rho}\delta_{\nu\sigma}
           +q\cdot (p+q)\delta_{\mu\rho}\delta_{\nu\sigma}
           +(\mu\leftrightarrow\nu)\biggr]. \tag$62$ $$
\bigskip
\noindent The linear term of the inner momentum $p_{\mu}$ gives rise
to the singularity of $1/\epsilon$ and this cancels out the vertex
factor $1/\alpha(d)$ in $d=4$ limit. Then we obtain a finite self-energy term,
which is independent of the gauge parameters. From gauge invariance,
this finite contribution is common to all the pieces of the invariant,
$F_{\mu\nu}F^{\mu\nu}$, so it
shifts the gauge coupling constant by the order of $1/N$. This can be
generalized to the other dimensionless coupling-constants, which are
included in the original lagrangian. On the contrary, we can show that
the effect on the self-energy of a scalar or of a fermion vanishes
at $d=4$ when the vertex is restricted to the one whose legs of
$h_{\mu\nu}$ are contracted by $P^{(2)}_{\mu\nu\lambda\sigma}$. \par
 Then the main quantum-effect of the gravitational fluctuations in four
dimension is the generation of the quadratic curvatures, which do
not exist in the original lagrangian, and the small shift of the dimensionless
coupling-constants in the original lagrangian. The magnitude of the shift
is finite but very small since it is the correction of the order of $O(1/N)$.
This shift is different from the case of the
topology changing effect due to the wormholes [13] or the the interactions
among many universes [14], where large shift of the parameters
is expected.
\bigskip
\noindent{\bf 6. Conclusion}
\bigskip
 We have investigated the quantum effects of the gravitation at
$4-\epsilon$ dimension, where the Einstein gravity is asymptotically safe
within the 1/N-expansion scheme. The perturbation in $d=4-\epsilon$
is well defined if we adopt the spin-3/2 field as N matter-fields, which are
included in the effective lagrangian as a loop-correction through a general
coordinate invariant regularization. In our approach,
the curvature squared terms are not introduced in the theory both at
$d=4-\epsilon$ and at $d=4$ to maintain the unitarity at any calculational
step. This was possible since the higher derivative terms are
not demanded even if we take the the limit of $\epsilon=0$ for the
calculated quantities in our scheme.
Our calculational procedure
is corresponding to performing the four-dimensional 1/N-expansion with
the constraint, $\int d^4x$(curvature squared terms)$=0$, at all stage
of the calculation.
However, our approach from the theory defined at $d=4-\epsilon$ would
be more tractable than the corresponding
four-dimensional calculation with constraints mentioned above. \par
 According to our formalism, we have examined the quantum effect of the
gravitation on the effective potential of a scalar field. In both
cases where the scalar has a non-minimal coupling with the gravity and
does not have, the effective potential including the graviton-loop
correction could not provide any non-trivial minimum which leads to the
spontaneous mass generation in $d=4-\epsilon$. In the limit of $\epsilon=0$
of the effective potential,
the logarithmic term of the scalar field vanishes leaving a term which
is proportional to the tree potential with the coefficient of order $O(1/N)$.
Then the quantum gravitational effect would be independent of a spontaneous
symmetry breaking for some symmetric theory within our calculational
scheme. \par
 On the other hand, the non-vanishing finite contribution of quantum
gravitational-fluctuation at $d=4$ can be generalized to the other field
operators of mass-dimension four. And this contribution is equivalent
to a small shift of the dimensionless parameters
if such operators exist in the original lagrangian. If they do not exist,
it provides new four-dimensional terms as a result of the
quantum-gravitational effect. For example, we have seen such terms,
$R_{\mu\nu}R^{\mu\nu}$ and $R^2$, above. But we should not use the newly
generated gravitational terms in the
perturbation, because they are the effective terms induced by the quantum
effect of the gravitation.
\bigskip
\noindent{\bf Acknowledgement:} The author thanks to the members of
high-energy group of Kyushu for useful discussions.

\bigskip
\bigskip
\noindent {\bf Appendix:}
Here we give the details of the calculation of the fermion-loop given
in section {\bf 3}.
The diagrams which contribute to the graviton propagator of momentum
$p$ are shown in Figs.3.a,b.
\bigskip
\centerline {\bf Fig.\ 3.a,b}
\bigskip
\noindent The vertices of fermion-graviton are given as follows,
\smallskip
$$\align V_{\mu\nu}^{(1)}&={1 \over 4}\delta_{\mu\nu}
                           [\gamma^{\lambda}(p+2q)_{\lambda}
                           +2mi] \\
        & \qquad \qquad\qquad-{1 \over 8}[\gamma_{\mu}(p+2q)_{\nu}+
                           \gamma_{\nu}(p+2q)_{\mu}] \tag$A1$
\endalign $$
\bigskip
\noindent for Fig.4a. And, for Fig.4b

$$\align V_{\mu\nu\lambda\sigma}^{(2)}&={1 \over 4}[
           \gamma\cdot q'(\delta_{\mu\nu}\delta_{\lambda\sigma}
            -2\delta_{\mu\lambda}\delta_{\nu\sigma})
            +{3 \over 2}(\gamma_{\sigma}q'_{\mu}\delta_{\lambda\nu}
            +\gamma_{\nu}q'_{\lambda}\delta_{\mu\sigma}) \\
           & \qquad\qquad\qquad -\gamma_{\nu}q'_{\mu}\delta_{\lambda\sigma}
            -\gamma_{\sigma}q'_{\lambda}\delta_{\mu\nu}]
            +i({\tilde {S_2}}+{\tilde S}_{2m}), \tag$A2$
\endalign $$
\bigskip
\noindent where
$$\align i{\tilde {S_2}}&={1 \over 16}\biggl \{
            -(\gamma_{\mu}\gamma\cdot p\gamma_{\lambda}
            +\gamma_{\lambda}\gamma\cdot p'\gamma_{\mu})
            \delta_{\nu\sigma}
            +4[(p+p')_{\mu}\gamma_{\sigma}\delta_{\nu\lambda} \\
            &\qquad\qquad +(p+p')_{\lambda}\gamma_{\nu}\delta_{\mu\sigma}]
            -2[(p+p')_{\mu}\gamma_{\nu}\delta_{\sigma\lambda}
            +(p+p')_{\lambda}\gamma_{\sigma}\delta_{\mu\nu}] \\
           &\qquad \qquad \qquad
          +\gamma\cdot (p+p')(2\delta_{\mu\nu}\delta_{\lambda\sigma}
                              -5\delta_{\mu\lambda}\delta_{\nu\sigma})
            \biggr \},  \tag$A3$ \\
        i{\tilde S}_{2m}&=i{m \over 4}
           (\delta_{\mu\nu}\delta_{\lambda\sigma}
            -2\delta_{\mu\lambda}\delta_{\nu\sigma}). \tag$A4$ $$
\endalign$$
\bigskip
\centerline {\bf Fig.\ 4.a,b}
\bigskip
\noindent Here the $\gamma$-matrices are the usual number matrices, and
$m$ is
the fermion-mass. \par
Then the self-energy corresponding to Fig.3.a,b are given as follows,

$$\align
  &\Pi_{\mu\nu\lambda\sigma}^{(a)}(p,m)=\int {{d^dq} \over {(2\pi)^d}}
    {{V_{\mu\nu\lambda\sigma}^{(a)}(q',p)} \over
         {(q^2+m^2+i\epsilon)([p+q]^2+m^2+i\epsilon)}}, \tag$A5$ \\
        &\Pi_{\mu\nu\lambda\sigma}^{(b)}(m)=\int {{d^dq} \over {(2\pi)^d}}
    {{V_{\mu\nu\lambda\sigma}^{(b)}(q,p)} \over
        {(q^2+m^2+i\epsilon)}}, \tag$A6$
\endalign $$
\bigskip
\noindent where $q'_{\mu}=q_{\mu}+p_{\mu}/2$, and

$$\align
    &V_{\mu\nu\lambda\sigma}^{(a)}(q',p)=tr\biggl [
        V_{\mu\nu}^{(1)}(\gamma\cdot q-im)
        V_{\lambda\sigma}^{(1)}(\gamma\cdot [p+q]-im)\biggr], \tag$A7$ \\
    &V_{\mu\nu\lambda\sigma}^{(b)}(q,p)=-tr\biggl [
        V_{\mu\nu\lambda\sigma}^{(2)}(\gamma\cdot q-im)\biggr]. \tag$A8$
\endalign $$
\bigskip
\noindent Exponentiating the denominators of Eqs.(A5,6) and integrating
over the momentum, they are written as

$$\align
     \Pi_{\mu\nu\lambda\sigma}^{(a)}(p,m)&
          -{{e^{i\pi d/4}} \over {(4\pi)^{d/2}}}
           \int_0^\infty d\alpha_1 d\alpha_2{1 \over {\alpha_+^{d/2}}}
           exp \biggl(i[\alpha_+m^2+
           {{\alpha_1\alpha_2} \over {\alpha_+}}p^2] \biggr) \\
        &\qquad\qquad\qquad
        V_{\mu\nu\lambda\sigma}^{(a)}(-i\partial_{z_1},-i\partial_{z_2})
          f^{(a)}(z_1,z_2)\big\vert_{z_1=z_2=0}, \tag$A9$ \\
        \Pi_{\mu\nu\lambda\sigma}^{(b)}(m)&=
          -i{{e^{i\pi d/4}} \over {(4\pi)^{d/2}}}
           \int_0^\infty d\alpha{1 \over {\alpha^{d/2}}} \\
        &\qquad\qquad\qquad exp(i\alpha m^2)
           V_{\mu\nu\lambda\sigma}^{(b)}(-i\partial_z)
          f^{(b)}(z)\big\vert_{z=0}, \tag$A10$
 \endalign $$
\bigskip
\noindent where $\alpha_+=\alpha_1+\alpha_2$, and

$$ \align &f^{(a)}(z_1,z_2)=exp\biggl(i[
                 {{\alpha_1-\alpha_2} \over {2\alpha_+}}p\cdot z_1
                 -{{z_1^2} \over {4\alpha_+}}+p\cdot z_2]\biggr), \tag$A11$ \\
          &f^{(b)}=exp(-i{1 \over {4\alpha}}z^2). \tag$A12$
\endalign $$
\bigskip
\noindent Here $z_1$ and $z_2$ are constant vectors.\par
 In $\Pi^{(a)}$, the following type of integral appears,
$$        I=\int_0^\infty d\alpha_1 d\alpha_2{1 \over {\alpha_+^k}}
           exp \biggl(i[\alpha_+m^2+
           {{\alpha_1\alpha_2} \over {\alpha_+}}p^2] \biggr) \tag$A13$ $$

where the exponent $k$ in $\alpha^k_+$ is either $1+d/2$ or $2+d/2$.
In any case, the integral $I$ can be rewritten as follows,

$$\align  &I=\int_0^1 d\alpha_1 d\alpha_2\delta(1-\alpha_+)I_k, \tag$A14$ \\
          &I_k=(-iA)^{k-2}\mathop{lim}_{\eta \to 0}
                \int_{\eta A}^{\infty} dx x^{1-k}e^{-x}, \tag$A15$
\endalign $$
\smallskip
\noindent where
$$ A=m^2+\alpha_1\alpha_2 p^2. \tag$A17$ $$

\noindent We obtain for $d=4-\epsilon$,

$$\align I_{2+d/2}&=(-i)^{d/2}\biggl\{\mathop{lim}_{\eta \to 0}
            \biggl[ {1 \over {2-\epsilon/2}}\eta^{-2+\epsilon/2}
            -{A \over {1-\epsilon/2}}\eta^{-1+\epsilon/2}\biggr] \\
          &\qquad\qquad\qquad
          \qquad\qquad\qquad +A^{d/2}\Gamma(-{d \over 2}) + O(\eta)
                     \biggr\}, \tag$A18$ \\
         I_{1+d/2}&=(-i)^{d/2-1}\biggl\{\mathop{lim}_{\eta \to 0}
            \biggl[ {1 \over {1-{\epsilon/2}}}\eta^{-1+\epsilon /2}\biggr] \\
          &\qquad\qquad\qquad
          \qquad\qquad\qquad +A^{d/2}\Gamma (1-{d \over 2}) + O(\eta)
                     \biggr\}. \tag$A19$
\endalign $$

\noindent Since $\Pi^{(a)}$ contains the terms $I_{2+d/2}$, $I_{1+d/2}$ and
$AI_{1+d/2}$, there are two types of divergence in $\Pi^{(a)}$,
$\eta^{-2+\epsilon/2}$ and $A\eta^{-1+\epsilon}$.\par
 In the case of $\Pi^{(b)}$, the following integral appears,

$$\align  {\tilde I}&=\int_0^{\infty} d\alpha \alpha^{1-k}
                      exp(i\alpha m^2) \tag$A20$ \\
          &=(-im^2)^{k-2}\mathop{lim}_{\eta \to 0}
                \int_{\eta m^2}^{\infty} dx x^{1-k}e^{-x}, \tag$A21$
\endalign $$
\smallskip
\noindent where $k$ is either $1+d/2$ or $2+d/2$ as in the previous
case. And
$m^2\eta^{-1+\epsilon}$ and $\eta^{-2+\epsilon}$ are divergent. In any case,
the divergences in $\Pi^{(a,b)}$ at $\eta=0$ could be regularized by the
Pauli-Villars method by requiring Eqs.(13,14). \par

\vfill\eject

\noindent {\bf References}\par
\bigskip
\noindent [1] K.S. Stelle, Phys. Rev., D16(1977)953; J. Julve and M.
Tonin, Nuovo Cimento, 46B(1978)137. The review is seen in I.Buchbinder,
S.Odintsov and I. Shapiro, Effective action in quantum gravity,
(IOP, Bristol, 1992).\par
\noindent [2] E. Tomboulis, Phys. Lett., 70B(1977)361.\par
\noindent [3] T. Lee and G. Wick, Nucl. Phys., B9(1969)209; R. Cutkosky,
P. Landshoff, D. Olive and J. Polkinghorn, Nucl. Phys., B12(1969)281;
D. Boulware and D. Gross, Nucl. Phys., B232(1984)1. \par
\noindent [4] L. Smolin, Nucl. Phys., B208(1982)439. \par
\noindent [5] L. Crane and L. Smolin, Nucl. Phys., B267(1986)714. \par
\noindent [6] D. Mckeon and T. Sherry, Phys. Rev., D15(1987)3854.\par
\noindent [7] R. Mann, L. Tarasov, D. Mckeon and T. Stelle, Nucl. Phys.
B311(1988)630.\par
\noindent [8] H. Kawai and M. Ninomiya, Nucl. Phys., B336(1990)115.\par
\noindent [9] J. Zin-Justin, Quantum field theory and critical phenomena,
(Oxford Scientific Pub. 1989) p.753. \par
\noindent [10] M.J. Perry, Nucl. Phys., B143(1978)114.\par
\noindent [11] L. Smolin, Nucl. Phys., B160(1979)253; A. Zee, Phys. Rev.,
D23(1981)858; S. Ichinose, Nucl. phys., B231(1984)335; K. Ghoroku,
Phys. Lett., 159B(1985)275. \par
\noindent [12] S. Coleman and E. Weinberg, Phys. Rev., D7(1973)1888.\par
\noindent [13] S. Coleman, Nucl. Phys., B310(1988)643. \par
\noindent [14] T. Banks, Nucl. Phys., B309(1988)493; S. Giddings and
A. Strominger, Nucl. Phys., B321(1989)481; K. Ghoroku, Class. Quantum
Grav., 8(1991)447.
\par
\vfill\eject
\noindent {\bf Figure Captions}
\bigskip
\noindent {\bf Fig.\ 1} (a) The loop-diagram contributing to the potential
of Eq.(50). The dashed line (the wavy lines) represents the scalar
field (graviton). (b) The diagram contributing to the potential of
Eq.(51). \par
\noindent {\bf Fig.\ 2} (a) The loop-diagram contributed to the potential of
Eq.(55), where the dashed line represents the scalar. (b) The
self-energy of the graviton. The solid lines represent the spin-3/2
fields. (c) The self-energy of a gauge boson, which is denoted by the
dotted line. The wavy line represents the graviton.\par
\noindent {\bf Fig.\ 3} The two types of the self-energy of the graviton
(the wavy lines) with the momentum $p$. The solid line represents
the spin-1/2 fermion.\par
\noindent {\bf Fig.\ 4} (a) The three-point and (b) the four-point vertex
of the fermion (the solid lines) and the graviton (the wavy lines).\par
\vfill\eject

\enddocument